\documentclass[preprint]{JASA}

\newcommand{\blue}[1] {\textcolor{blue}{#1}}

\begin{document}

\title[]{Blinded Evaluation of Oceanic Sound Source Locations via Sequential Bound Estimation}


\author{John L. Spiesberger}
\email{john.spiesberger@gmail.com}
\affiliation{Dept. of Earth and Environmental Science, U. of Pennsylvania, Philadelphia, PA 19104, USA}
\author{Julien  Bonnel}
\email{jubonnel@ucsd.edu}
\affiliation{Marine Physical Laboratory,  Scripps Institution of Oceanography, UC San Diego, 9500 Gilman
  Drive \#0238, La Jolla, CA 92093-0238 USA}






\email{john.spiesberger@gmail.com}




\begin{abstract}
 A non-linear, non-Bayesian method called sequential bound estimation (SBE) derived 100\%  confidence
  intervals of location (CIL) for 219 explosions in
  the ocean from measurements of their time differences of arrivals (TDOA) among five widely-spaced time-unsynchronized receivers on the ocean bottom.
  The explosion's locations were measured with the global  positioning system.
  A blind evaluation revealed all  219 explosions  were within their CIL. 
  The probability this could happen
  by chance is $4  \times 10^{-116}$. The explosions were  detonated in shallow water on the eastern continental shelf of the U.S. 
  over the so-called New England Mud Patch.

\end{abstract}


\maketitle



\section{\label{sec:intro} Introduction}

Sources of electromagnetic (EM) and acoustic waves are routinely located 
from measurements of their time-differences-of-arrival (TDOA) between
widely separated receivers
in space, air, water, and solid Earth.  The accuracy
of location is affected by errors in the TDOA, receiver's locations and clocks, speed of acoustic or EM waves,
and the location method
itself. Reliable methods must account for these errors.   Here, we conduct a blind evaluation of Sequential bound estimation (SBE)
using sounds from hundreds of explosions detonated in shallow water.

 SBE is a non-linear non-Bayesian estimator designed to explicitly estimate all phenomena affecting location \citep{prob_distr, sbe}.
It was developed to provide reliable 100\% confidence intervals (CI) for all the relevant variables affecting location so it might be possible
to yield reliable confidence intervals of  location (CIL).  It is  related to interval methods, also yielding
100\% CIL though with a smaller number of variables affecting location to make the computations practical \cite{MUSTAFA2018160,Codres2022GuaranteedS}.
SBE is also related to efficient non-linear Bayesian approaches that infer 
probability distributions for all relevant variables \cite{warner2016bayesian,gehrmann2026,snyder_whaledo}.

Past evaluation of SBE yielded CIL always containing the true locations of sounds from simulations \cite{cse_eval,ishmael} and two experiments. The first experiment generated
the 100\% CIL for one bird call \cite{sbe}.  The second experiment yielded 100\% CIL for the locations of a submarine from Naval Air's multistatic active sonar, where 100\% CI where
also generated for locations
of drifting receivers, the speeds of sound, and the receiver's clocks \cite{sbir_phase2}.   The study here differs because its evaluation is blinded,
deals with many more locations, and is unclassified, making its results available to anyone, unlike the classified Naval Air evaluation.

The data used for SBE here  were collected in 2022 as part of the Seabed Characterization
Experiment (SBCEX) \cite{wilson2020seabed,wilson2022guest}.  Hundreds of explosions were detonated at locations measured with the global positioning system (GPS) and recorded on
twenty time-unsynchronized receivers light enough to toss into
the water by hand, thus their name: TOSSITs \cite{zitterbart2022tossit, bonnel2022geoacoustic}. In practice, the TOSSITs are gently lowered to  water level and released.
They incorporate SoundTrap recorders
manufactured by Ocean Instruments \cite{OceanInstruments2021}.  In this paper, the TDOA were derived from four or five receivers, and inputted to SBE.
The reliability of SBE was assessed through a formal blind evaluation. To do so, the evaluation was performed by the two co-authors here,
with the true source locations known only by one. In practice, Dr. Spiesberger ran SBE and generated  the 100\% CIL without knowledge of any true locations. These CIL  were
sent to Dr. Bonnel who had access to true source locations, and checked if the CIL were correct.

This  paper is organized as follows.  Some methods for obtaining reliable locations with TDOA are discussed in Sec. \ref{sec:tdoa_loc}.
More information about SBE's approach appears in Sec. \ref{sec:sbe_method}.  The experiment is described in Sec. \ref{sec:experiment}.
Sec. \ref{sec:data_assoc} discusses  the method for associating sounds from the experiment with measured arrival times at the receivers.
Parameters used for SBE's solution are described in Sec. \ref{sec:sbe_prior}. Results of SBE's blind evaluation appear in
Sec. \ref{sec:results}.  A discussion of the  results and conclusions appears in Secs. \ref{sec:discussion} and \ref{sec:conclusion}.

\section{\label{sec:tdoa_loc} Reliable Locations via TDOA}

For two- and three-dimensional models of location (2D and 3D), \citet{tyrell} and \citet{schmidt} showed it necessary to have at
least four and five receivers respectively
to yield a guaranteed mathematical unambiguous solution for
location from TDOA in the absence of errors.  The
TDOA, $\tau$, can be transformed into the calling animal's difference of distance, $\delta d$, from a pair of receivers using,
$ \delta d = c \tau$, when $c$ is a known speed of sound. The locus of points in space sharing this difference in distance 
for a pair of receivers defines the hyperboloid, in three spatial dimensions, and a hyperbola, in two dimensions
\citep{merriam_webster}. Adding receivers
adds hyperbolas and their
intersections coincide at a single point with four and five receivers for 2D and 3D models respectively.  This approach yields
reliable locations for small enough errors of the clocks,  receiver's locations, and the speed of sound.

Reliable CIL might be obtained for very small errors when jointly estimating the sound  speed and wind fields via tomography with
a linearized approximation \cite{pass_loc}. \citet{schmidt} noted problems with deriving CIL via the long range navigation system (LORAN) using
a linear approximation between location and the TDOA. For many cases of interest, the linearized approximation is invalid
and non-linear methods must be employed \cite{prob_dens,warner_2017}.
Hyperbolas are not sufficiently accurate enough to yield accurate CIL when the speed of sound
is not very nearly  spatially homogeneous \cite{prob_dens}, motivating a different geometrical shape called an isodiachron for deriving location
\cite{isodiachrons}.
Even when the speed of sound is a known constant, temporal interference between acoustic paths at a receiver
may  render the hyperbola invalid for purposes of location when the sources
are near the receivers and at least is one is near the ocean's surface \cite{Spiesberger2025SlowingTS,spies_terray_2026}. Locations derived with
hyperbolas,  or equivalently, any methods converting TDOA to  difference in distance with $\delta d = c \tau$, for constant $c$, are also invalid
when  the vertical  coordinate of a source is removed from
 the problem, leaving only a solution for its horizontal location. Removing the vertical coordinate
 yields so-called ``2D  black  holes'', yielding significant errors unless accounted for \cite{2d_black_holes}.
 A narrated tutorial of 2D black holes can be found in the Supplementary material in \citet{cse_eval}.
Errors in the receiver's clocks can generate large errors in locations \cite{warner_2017,cse_eval,ishmael}, so it is best to jointly solve for clock
errors as well.

\section{\label{sec:sbe_method} TDOA location via SBE}

SBE is a non-linear non-Bayesian method for locating sounds via TDOA designed to account for all phenomena affecting location
\citep{prob_distr, sbe, patent_sbe2, patent_sbe3, patent_sbe4, patent_sbe5,patent_clock_sync1, patent_clock_sync2, 2d_black_holes}.
 Although SBE explicitly estimates errors in the receiver's clocks, it is aided by an algebraic
 solution useful for bounding their errors from the TDOA themselves \cite{patent_clock_sync1,patent_clock_sync2}. It accounts for spatial inhomogeneities
 of sound speed via isodiachrons \cite{isodiachrons}, effects of temporal interference, and the 2D approximation (Sec. \ref{sec:tdoa_loc}).
 SBE is designed to yield
 100\% confidence intervals, as opposed to non-linear Bayesian approaches
yielding confidence intervals less than 100\%. SBE's CIL are independent of the shapes of  the joint prior
probability distributions of error required by all  Bayesian  techniques. Since the prior distributions of error are almost always unknown,
unreliable results can be obtained with Bayesian approaches if  there are insufficient data to reliably estimate their posterior distributions.
SBE's independence of these prior distributions is its leading cause of reliability.  Instead, it is necessary to set prior bounds for all the 
variables affecting location. Since the smallest prior bounds are almost always unknown, their limits are set large enough so  there is
no chance they are too small. This is always possible, yet small  CIL are often obtained with SBE because these bounds are updated in the light of data.
It is beyond the scope of this paper to further describe SBE. The references contain all this information.

\section{\label{sec:experiment} Experiment}

The data considered in this paper were collected in 2022 as part of the Seabed Characterization
Experiment (SBCEX). SBCEX is a multi-year multi-institutional at-sea research effort that took
place on the New England Shelf, with a major focus on the New England Mud Patch (NEMP), 95 km south of
Martha's Vineyard (MA, USA). While the main objective of SBCEX was to study the interaction of sound with fine-grained
sediments \cite{wilson2020seabed,wilson2022guest}, its scope was extended to evaluate the performance of algorithms and  
develop new techniques in signal processing, source localization, and geophysical inversion.

The data considered here are recordings of so-called signal underwater sound (SUS) charges (Sec.~\ref{sec:sus}) from five passive acoustic moorings called
TOSSITs (Sec.~\ref{sec:tossits}), all deployed within the NEMP. The seabed in the experimental area has been extensively
studied through a high-resolution seismic survey \cite{goff2019stratigraphic}, geophysical and acoustic cores
\cite{Chaytor2021,ballard2024direct,lee2025investigation}, as well as multiple geoacoustic inversion studies, e.g.
\cite{bonnel2024trans, knobles2024feature, potty2025estimation}. The bathymetry of the experimental area is shown
in Fig.~\ref{fig:bathy_rec}.

The SUS-TOSSIT data under examination were previously used for deep-learning based geoacoustic inversion \cite{vardi2024end, vardi2024estimation},
but not for source localization.

\subsection{\label{sec:sus} Sound Sources}

A total of 219 SUS charges were detonated between 13:00 on 15 May  and 21:30 on 18 May 2022 GMT.
They were programmed to explode at depths of either 18 or 91 m. Deployed from the stern of a  ship,  their
sounds were recorded with a low-sensitivity hydrophone also towed from the
ship's stern at about 50 m depth. Detonation times were recorded by a GPS-synchronized IRIG-B microphone, and the GPS positions
were derived from a
handheld GPS unit at the stern of the ship.  Most of the GPS locations were recorded
while the ship was nearly stationary, and a few others while underway.


\subsection{\label{sec:tossits} Receivers}

During SBCEX22, a network of 20 TOSSITs, was deployed on the NEMP to record
the SUS explosions. A TOSSIT is a low-cost, ropeless, hand-deployable mooring with a single autonomous acoustic
recorder (SoundTrap ST300STD from Ocean Instruments, NZ) located $\simeq 1$~m above the seabed
\cite{zitterbart2022tossit, bonnel2022geoacoustic}.

\begin{figure}[ht] 
\centerline{\includegraphics[width=6in]{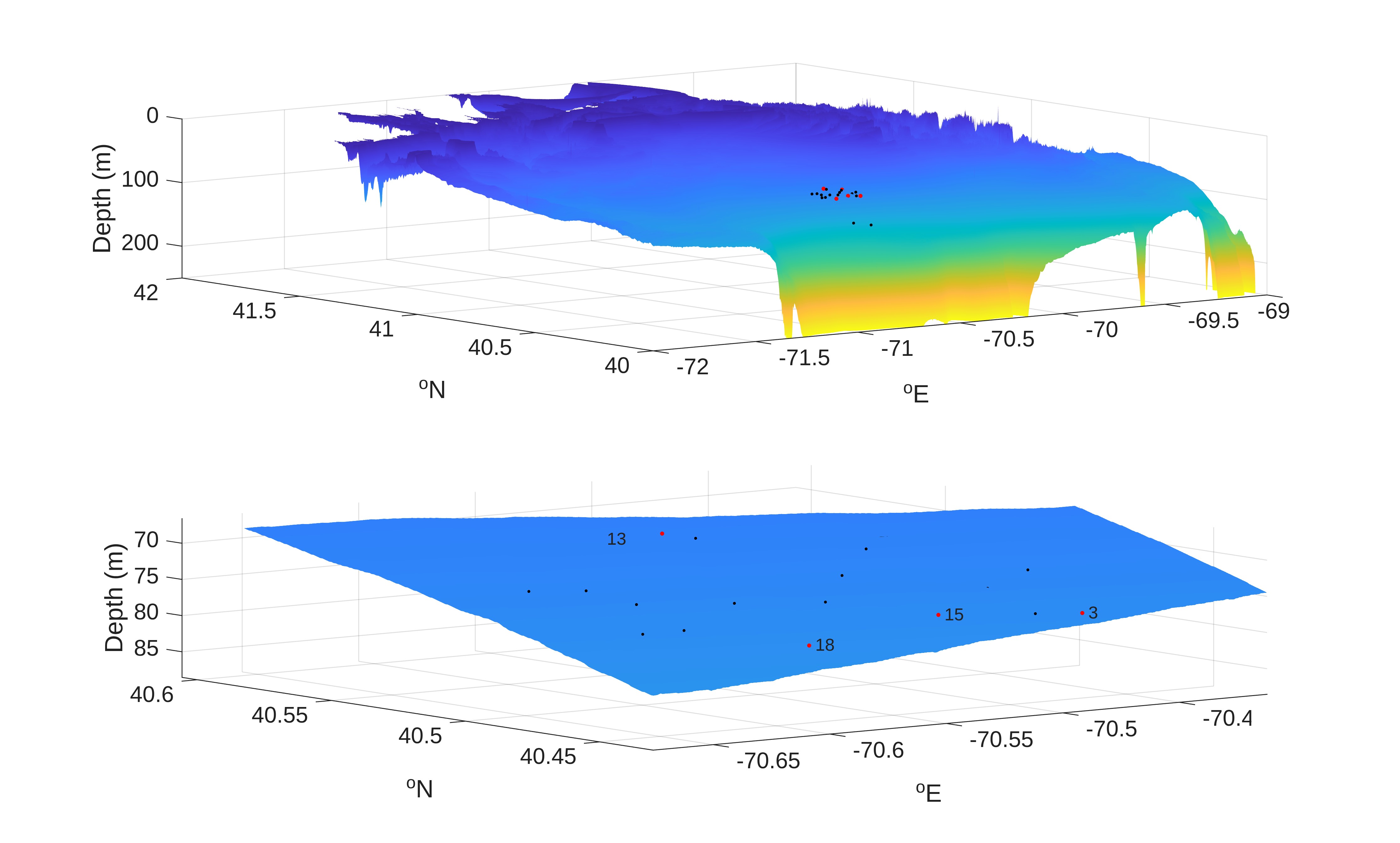}}
  \caption{\label{fig:bathy_rec} {\bf Top:} Bathymetry on the New England continental shelf near a region with a muddy bottom \cite{Chaytor2021}
  with locations of twenty receivers on top (black). The red dots are the five receivers
  whose data are used in this paper. {\bf Bottom}: Blow up with Bonnel receiver numbering.}
\end{figure}

In this study, a subset of 5 TOSSITs is considered. Their locations, as well as the local water depth at
their locations, are given in Table~\ref{tab:rec_locs}. The receivers are separated by about 6 km. The
experimental setup is illustrated in Fig.~\ref{fig:bathy_rec}: each of the 20 TOSSITs is represented
by a black dot, while the 5 specific TOSSITs considered here are highlighted in red.

\begin{table}[ht]
\caption{\label{tab:rec_locs}Approximate locations and depths of five receivers from Fig. \ref{fig:bathy_rec}.}
\begin{ruledtabular}
  \begin{tabular}{lll}
    RECEIVER&LONGITUDE/LATITUDE&DEPTH\\
    & ($\mbox{}^{\circ}$E , $\mbox{}^{\circ}$N))& (m)\\
    \hline
    3&(-70.4913 , 40.4303)&76.5\\
    7&(-70.5012, 40.5000)&70.5\\
    13&(-70.5673, 40.5210)&68.5\\
    15&(-70.5335 40.4473)&76.5\\
    18&(-70.5965, 40.4408)&78.5\\
\end{tabular}
\end{ruledtabular}
\end{table}

\section{\label{sec:data_assoc} Data Association}

For each explosion, it is necessary to identify the same explosive event on each receiver to derive the TDOAs. 
This is done automatically
by cross-correlating the temporal
pattern of a receiver's detected explosion times with the temporal pattern of detonation times.
Every explosion yields very high  signal-to-noise ratios  (SNR's), so the one-to-one
linking is straightforward.  Afterwards, each linking
is checked visually to make sure the pattern matchings are correct.

\section{\label{sec:sbe_prior} Prior Bounds for SBE}

For SBE to provide 100\% CIL for the source's locations, it is required to provide 100\% prior bounds for all parameters affecting localization.
  This is discussed here for the source and receiver locations (Sec.~\ref{sec:sus_prior_locs}), sound speed field (Sec.~\ref{sec:c_field}), receiver's
  clocks (Sec.~\ref{sec:clocks}) and TDOAs (Sec.~\ref{sec:tdoa_bnds}).

\subsection{\label{sec:sus_prior_locs} Receiver and SUS Locations}

The horizontal locations of receivers (Table \ref{tab:rec_locs}), measured with a handheld GPS receiver during the experiment, are assigned 100\% confidence intervals (CI) of $\pm 125$ m.
Fifty meters is due to error in the GPS measurements and their uncertain landing spots on the sea bottom. The remaining seventy-five meters is caused by moving centers
of the prior bounds away from the GPS measurements, then adding extra uncertainty to be 100\% sure the receiver is inside. We could have used smaller bounds but wanted
to determine if the TDOA data contained enough information to reduce them via SBE.

Received SNRs from the SUS charges exceed 40 dB, and in many cases, 60 dB,  thus can be detected at hundreds of kilometers.  The SNR of the first arrival is based on the
ratio of the peak-amplitude of the first arrival divided by the standard deviation of the noise,  evaluated in the band from 500 to 600 Hz (e.g. Fig. \ref{fig:first_arrivals}A2,B2).
The prior locations of the SUS charges
are set to a smaller region within [-70.8 -70.4] $\mbox{}^{\circ}$E and [40.14 to 40.55] $\mbox{}^{\circ}$N, encompassing the whole experimental area and large enough to  contain their
true locations. These bounds contain 33.9 km of longitude and 65.6 km of latitude.
The smaller region reduces computations needed to model the sound speed field.  Consequences of adopting
smaller bounds are discussed later (Sec. \ref{sec:sus_prior}).

\subsection{\label{sec:c_field} Sound Speed Field}

\begin{figure}[ht] 
  \centerline{\includegraphics[width=3.5in]{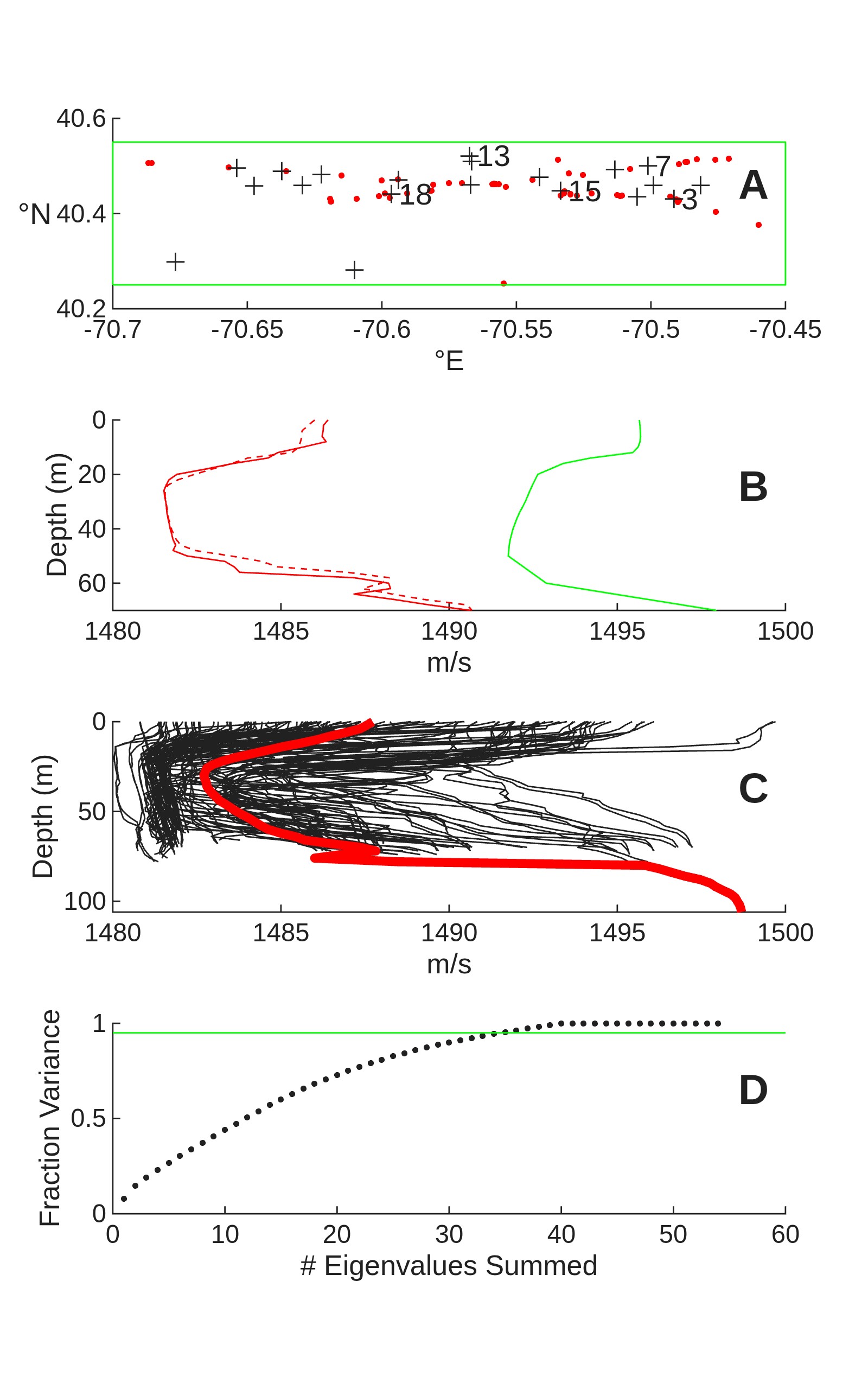}} 
\caption{\label{fig:ctd_info}  {\bf A}: Locations of fifty seven CTD  stations (red) near receivers (+). Numbered receivers used
  in this paper. {\bf B}: Speed of sound derived from downcast/upcast of one CTD in few minutes (red) compared with HYCOM model (green)
  from $1/12\mbox{}^{\circ}$ global analysis from 41-layer model at 3-hr temporal resolution \cite{Metzger2017,hycom_gofs31}. HYCOM model \cite{hycom_gofs31}, interpolated to CTD location,  is 2.2 hours
  earlier than CTD. {\bf C}: Sound speeds derived from 114 CTD profiles (57 downcast plus 57 upcasts) (black) and their mean (red). Data courtesy Bill Hodgkiss, Scripps Institution of
  Oceanography.
  {\bf D}: Fractional  variance
  of sound speeds as function of number of eigenvalues summed from EOF decomposition. Green is 95\% of total variance.
}
\end{figure}

CTD data collected between 12 and 26 May 2022 reveal a field of sound speed varying  by tens of meters per second over distances less than 10 km,
likely due to  energetic internal waves
(Fig. \ref{fig:ctd_info}A,C). Sound speeds derived from one downcast/upcast at $-70.4928 \ \mbox{}^{\circ}$E and
$40.4355 \ \mbox{}^{\circ}$N on 17 May 11:12:15 GMT 2022 are about 10 m/s slower than
derived from the HYCOM data at the same location and 2 hr 12 min  earlier  (Fig. \ref{fig:ctd_info}B). The HYCOM bias in shallow water is typical \cite{CAROLINACASTILLOTRUJILLO2023103126},
and is thus unused for our study.
Neighboring CTD stations reveal very different sound speeds with  depth, and thus are aliased. Furthermore, empirical orthogonal functions (EOF)
of speeds as a function of depth yield a whitish spectrum of eigenvalues  (Fig. \ref{fig:ctd_info}D), further
complicating derivation of prior bounds of sound speed.

Without a reliable data-assimilative model for the CTD data, ad-hoc 3D databases of sound speed
are constructed by gridding the geographic region within [-70.9 , -70.25] $\mbox{}^{\circ}$E and
[40.05 , 40.75] $\mbox{}^{\circ}$N at intervals between 5000 and 10000 m. A sound-speed EOF is randomly assigned to be the speed of sound with depth
at each geographic grid point.  In between grid points, the speed of sound is derived from a 2D linear  interpolation at any depth.

\begin{figure}[ht] 
  \centerline{\includegraphics[width=6in]{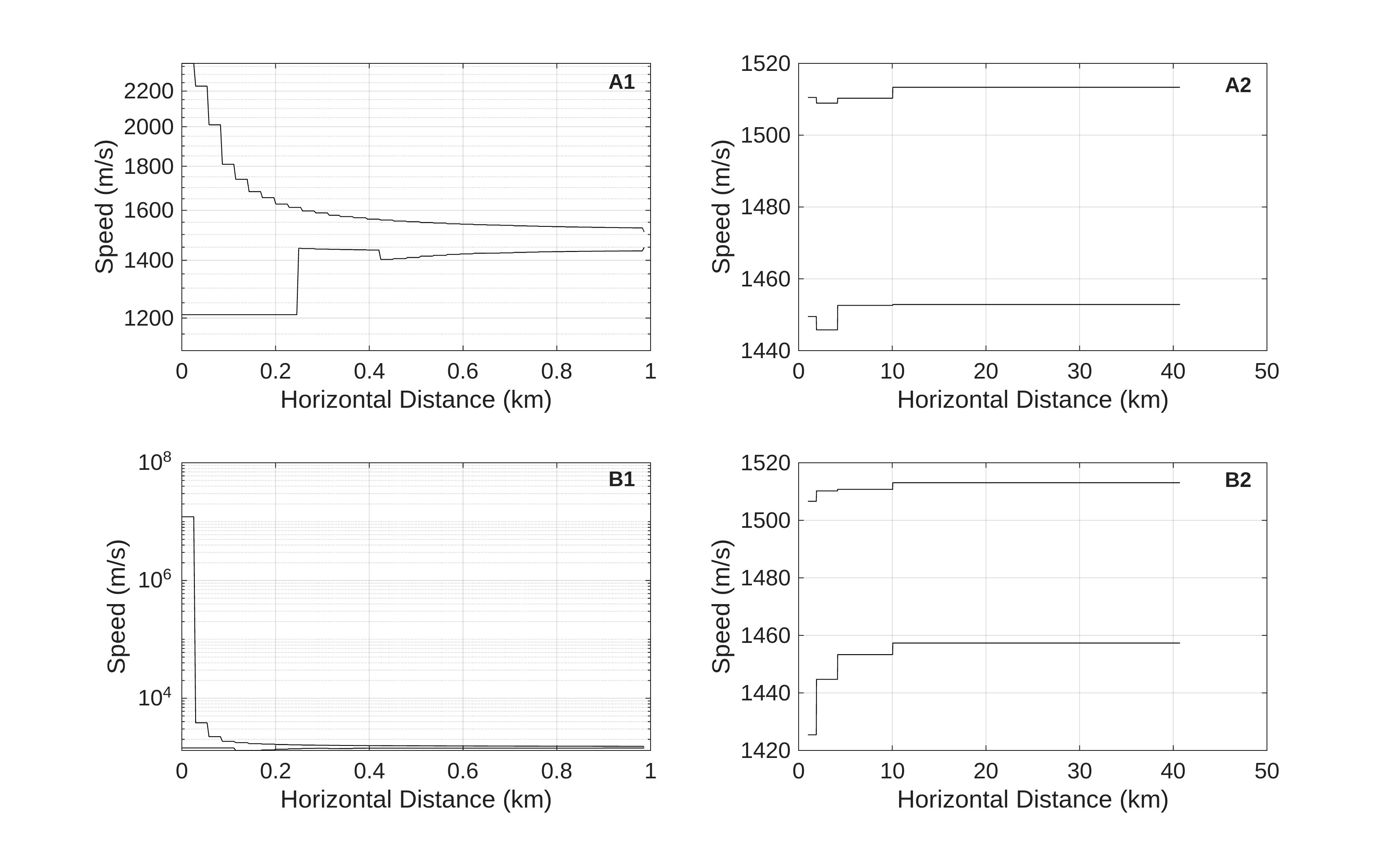}}
\caption{\label{fig:c3d_bnds_T03}  Bounds of $c_{3d}$ as function of horizontal distance from Bonnel receiver 3
  (Fig. \ref{fig:bathy_rec}) at depths of 10 and 50 m in top and bottom rows respectively.  Bounds near receiver
  on logarithmic axes with differing limits  because of widely different values for upper bound. Axis limits same
  for panels on right. Panel {\bf B1} has minimum of 1295.3 m/s.}
\end{figure}

SBE requires 100\% confidence intervals of the so-called 3D effective speed, $c_{3d}$, field, 
  \begin{equation}
c_{3d} \equiv \sqrt{(d/t)^2 + (\delta z)^2}\ , \label{eq:c3d}
  \end{equation}
  were $d$ is the  horizontal geodesic distance between the receiver and some longitude and latitude, $t$ is the propagation time of the first arriving acoustic path
  between a 3D grid point
  and the receiver, and $\delta z$ is the difference in depth between
  the 3D grid point and receiver. Because each receiver is surrounded by different bottom depths, sound speed fields are computed for each receiver separately as a function
  of  geographic location  and depth. Times of first arriving paths are computed with rays \cite{bowlin_ray} and the eigenray finder described by \citet{Spiesberger1994}.  Effects of horizontal refraction are neglected.  To mimic effects
  of unresolved temporal interference, the waveform along each  path is modeled to be a sine wave at frequency 550 Hz,  with Gaussian envelope of width 0.01 s.
  Uncertainties of the sound speed field are modeled by computing 5000 eigenray realizations for each grid point, where each realization randomly perturbs the
  eigenray time and amplitude by a few milliseconds and dB respectively. The perturbation of amplitude is due to uncertainty of the  sound speed field, 
  reflections from the wavy sea-surface, and uncertainties of the reflection coefficient at the ocean bottom.   Variations of eigenray time due to waves of about 1 m amplitude
  are derived from
  the ray's inclination at the surface.  The minimum and maximum values of $t$ are computed from the realizations to compute 100\% bounds of $c_{3d}$ at each grid point.
  Near  caustics, amplitudes of eigenrays are derived from theory [Eqs. (6.1.14), (6.1.19), and (6.1.27), \citet{brekh_godin}]. The 3D database of $c_{3d}$ is collapsed into a
  2D field by finding the min/max values of the $c_{3d}$ at any horizontal distance and depth  among all  bearing angles from the receiver. Ray parameters were halved until
  reaching convergence. Computations of the $c_{3d}$  bounds are carried out for nine 3D databases of sound speed, where convergence of the $c_{3d}$ bounds was realized
  (Fig. \ref{fig:c3d_bnds_T03}). An additional 20 m/s was added  to the $c_{3d}$ bounds to be sure they bounded the unknown $c_{3d}$ values.

Near the receivers, temporal interference causes the first arriving path come in sooner or later than the propagation time along an eigenray, leading to
$c_{3d}$ values significantly less or greater than the {\it in-situ} speed. These sub- and supersonic energy-bearing
wavepackets are discussed elsewhere \cite{Spiesberger2025SlowingTS,spies_terray_2026}. The large excursions of  the $c_{3d}$ from {\it in-situ} values
near a receiver, e.g. 1000 to $10^7$ m/s,
is a primary reason
for modeling its bounds as a function of horizontal distance.

\subsection{\label{sec:clocks} Clocks}

No measurements were made of clock errors at the start or end of the experiment.  To  make sure their prior bounds are large  enough,
they are set to $\pm 2 \times 10^4$ s relative to receiver 15's clock, the so-called ``reference receiver'', whose error is defined to be zero throughout
the experiment.  
The published fractional frequency error of the soundtrap ST300 is $2 \times 10^{-5}$ \cite{OceanInstruments2021}.  Thus, the fractional frequency
error of any clock relative to receiver 15's clock is twice as  large, namely 
$4 \times 10^{-5}$. 
After one day, the maximum
relative clock error is $\pm 4 \times 10^{-5}  \times 86400 \ \mbox{s} =  \pm 3.5$ s.

\subsection{\label{sec:tdoa_bnds} TDOA}

\begin{figure}[ht!] 
  \centerline{\includegraphics[width=5in]{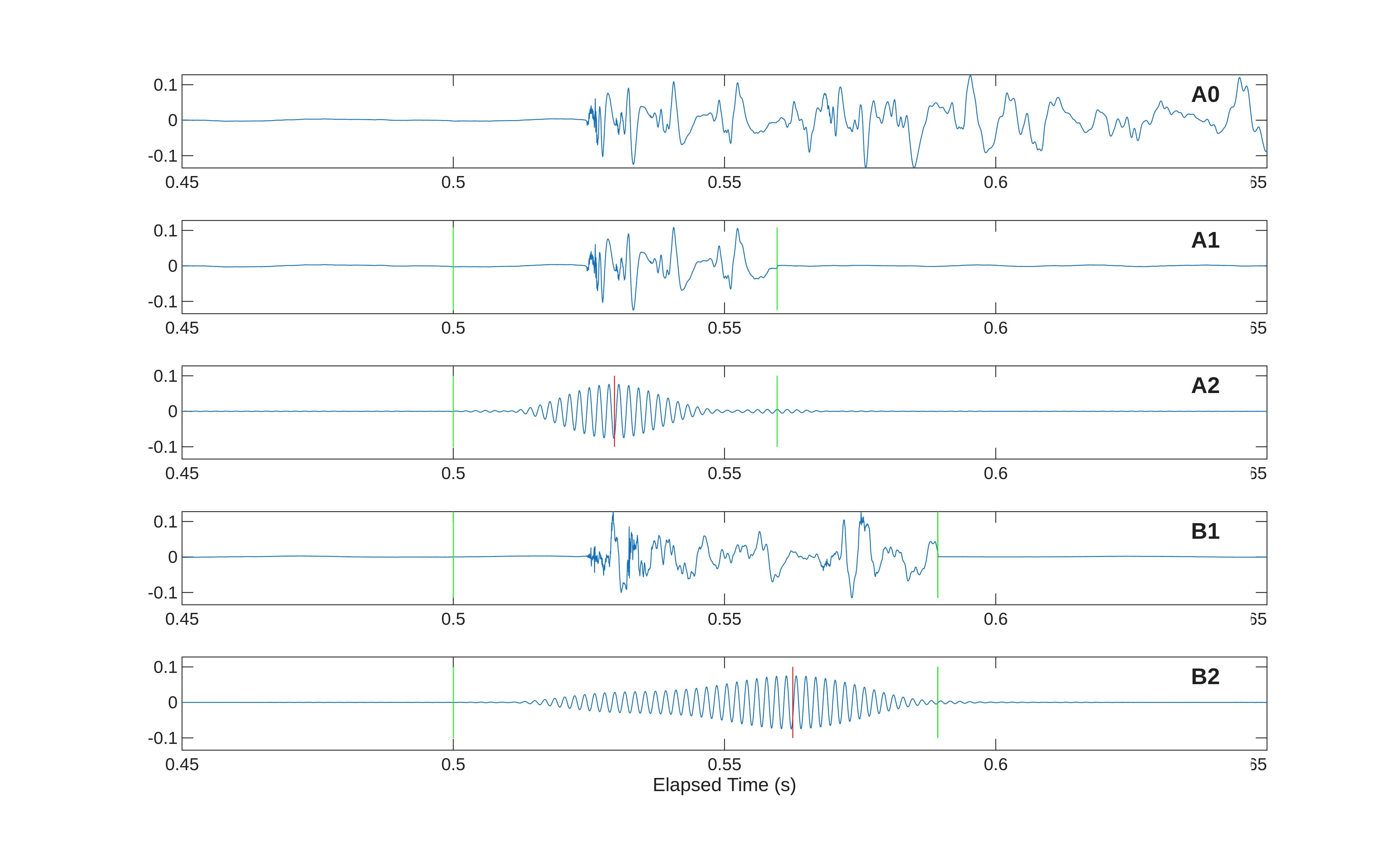}}
\caption{\label{fig:first_arrivals} Sounds from two SUS charges, A and B.  {\bf A0}: Recorded data from one explosion. {\bf A1}: Same as A0 except data following green window
  are replaced by ocean noise  proceeding green window to create time-limited energy near first arrival. {\bf A2}: Data in A1 multiplied by triangular taper reaching
  zero at green lines and peak  in center, followed by bandpassing between 500 and 600 Hz  (Sec. \ref{sec:tdoa_bnds}). {\bf B1} and {\bf B2}: same as {\bf A1} and {\bf A2} except
  for different explosion.
  Red lines indicate elapsed time corresponding to  maximum of absolute value of Hilbert transform.  Note temporal interference of at least
  two multipath in {\bf B2}  but not in {\bf A2}.
}
\end{figure}

TDOA  are estimated by finding the lag corresponding to the largest peak of
the cross-correlations of spectrograms, each  centered about the first
arriving multipath propagating through the water. Data are  prepared for the spectrogram as follows. First, explosive events are detected by seeking
peaks exceeding 30 dB; a threshold exceeded by all the explosions (Fig. \ref{fig:first_arrivals}A0).
Second, a boxcar time window is applied to the data to remove energy arriving after the first arriving energy (green lines, Fig. \ref{fig:first_arrivals}A1).
Then ocean noise preceding the boxcar window is added
to times following this window (Fig. \ref{fig:first_arrivals}A1). Third, the data in the window are multiplied by a triangular taper with peak in the middle
and dropping to zero at its boundaries.  These are bandpassed from 500 to 600 Hz to suppress
any early arriving energy traveling through the seabed.
(Fig. \ref{fig:first_arrivals}A2). Windowed data from another explosion appear in Fig. \ref{fig:first_arrivals}B1,B2.
The arrival time of the energy usually looks like one temporally-resolved arrival with half-height
duration  $\sim 0.01$ s (Fig. \ref{fig:first_arrivals}A2). However,  sometimes temporal interference with other multipath shifts the peak time of arrival
by $\sim 0.05$ s or more
(Fig. \ref{fig:first_arrivals}B2).

There are three contributions to uncertainties of the TDOA presented to SBE. The smallest is due to the SNR
and bandwidth of the spectrogram data.  This is usually a small effect with uncertainty of about $\sim 0.01$ s.  The second contribution is due to
interference between the first arriving multipath and later arriving energy (Fig. \ref{fig:first_arrivals}B2).
All 219 explosive receptions were inspected to estimate its maximum effect, and was about $\pm 0.1$ s, though most cases were much less.
The largest contribution is due to uncertainty of clock  timing errors in  OceanInstruments receivers.
The analog-to-digital (A/D) time stamps
of samples across audio file  boundaries have errors of several seconds (\blue{Appendix} \ref{sec:app1}).  There are two causes of these errors.
The largest effect can be corrected from data in the log files  outputted by the soundtraps (\blue{Appendix} \ref{sec:app1}). There may be another
error of $\pm 0.5$ s across audio file boundaries, but this cannot be corrected from data in the log files.  Having corrected the
larger time jumps of several seconds, we therefore assigned all TDOAs to have error of $\pm 0.5$ s, because the effects due to SNR, bandwidth, and temporal  interference
are much less. Explanations of these A/D timing errors
are unavailable.

\section{\label{sec:results} Results}

\begin{figure}[ht] 
  \centerline{\includegraphics[width=\linewidth]{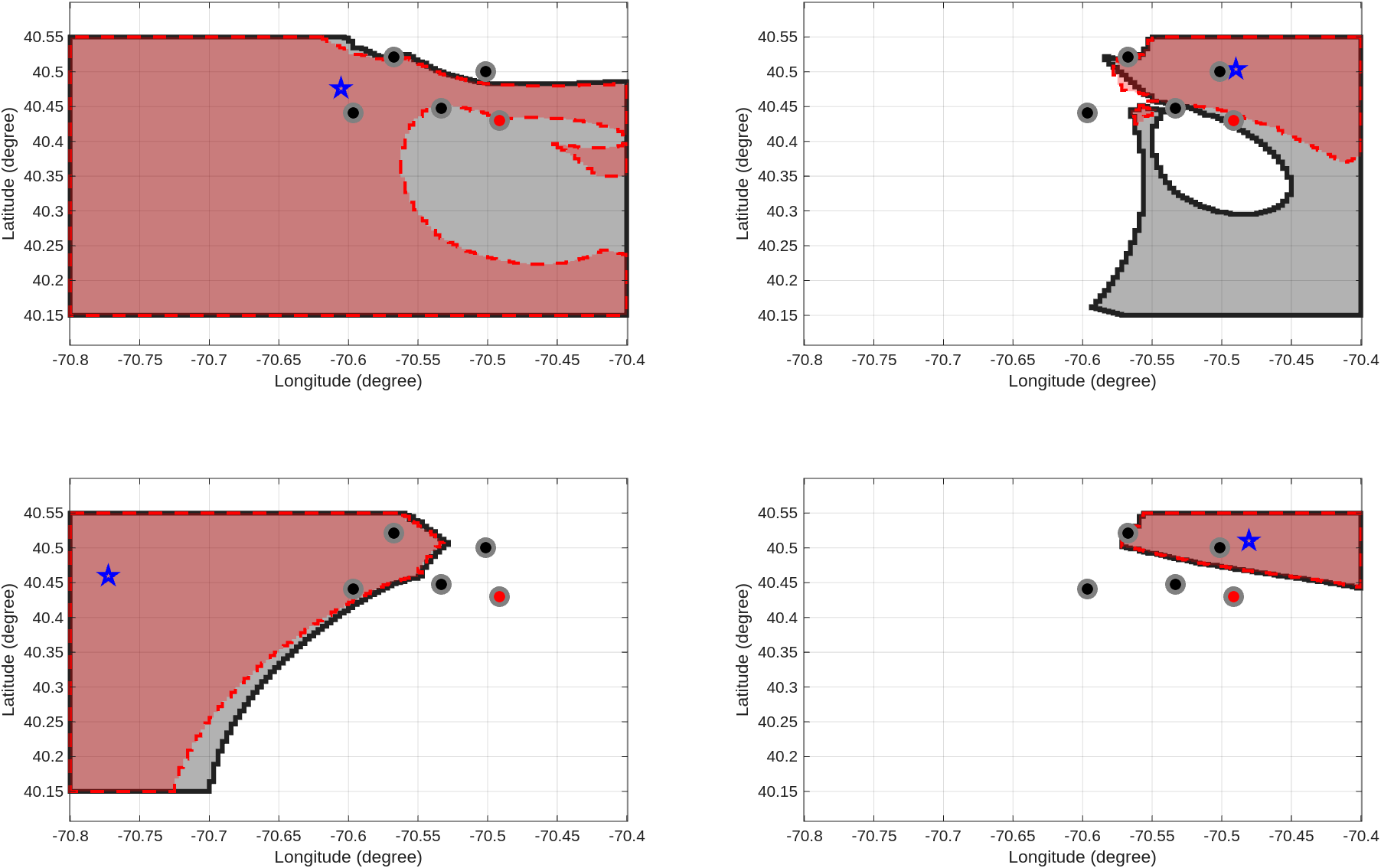}}
\caption{\label{fig:fig_for_article_draft}  100\% CIL derived with SBE for four SUS-charge explosions at GPS locations (blue stars)
  using four receivers  (grey-black circles) and then with addition of fifth receiver (grey-red circle). CIL  derived with five receivers (red-shading)
  are significantly smaller for explosions on top row, but not on bottom. Four receiver's CIL are partially covered by CIL from five-receiver red-shaded CIL.
}
\end{figure}

Data from 219 explosions were assimilated by SBE,  yielding 100\% CIL for their locations, the
receiver's locations, relative clock  errors, and the 3D  effective speeds between each explosion
and each receiver. Each CIL contained the explosion's GPS's location.
Fig. \ref{fig:fig_for_article_draft} shows the CIL for four explosions derived with four and five  receivers. 

\begin{figure}[ht!] 
   \centerline{\includegraphics[width=4.in]{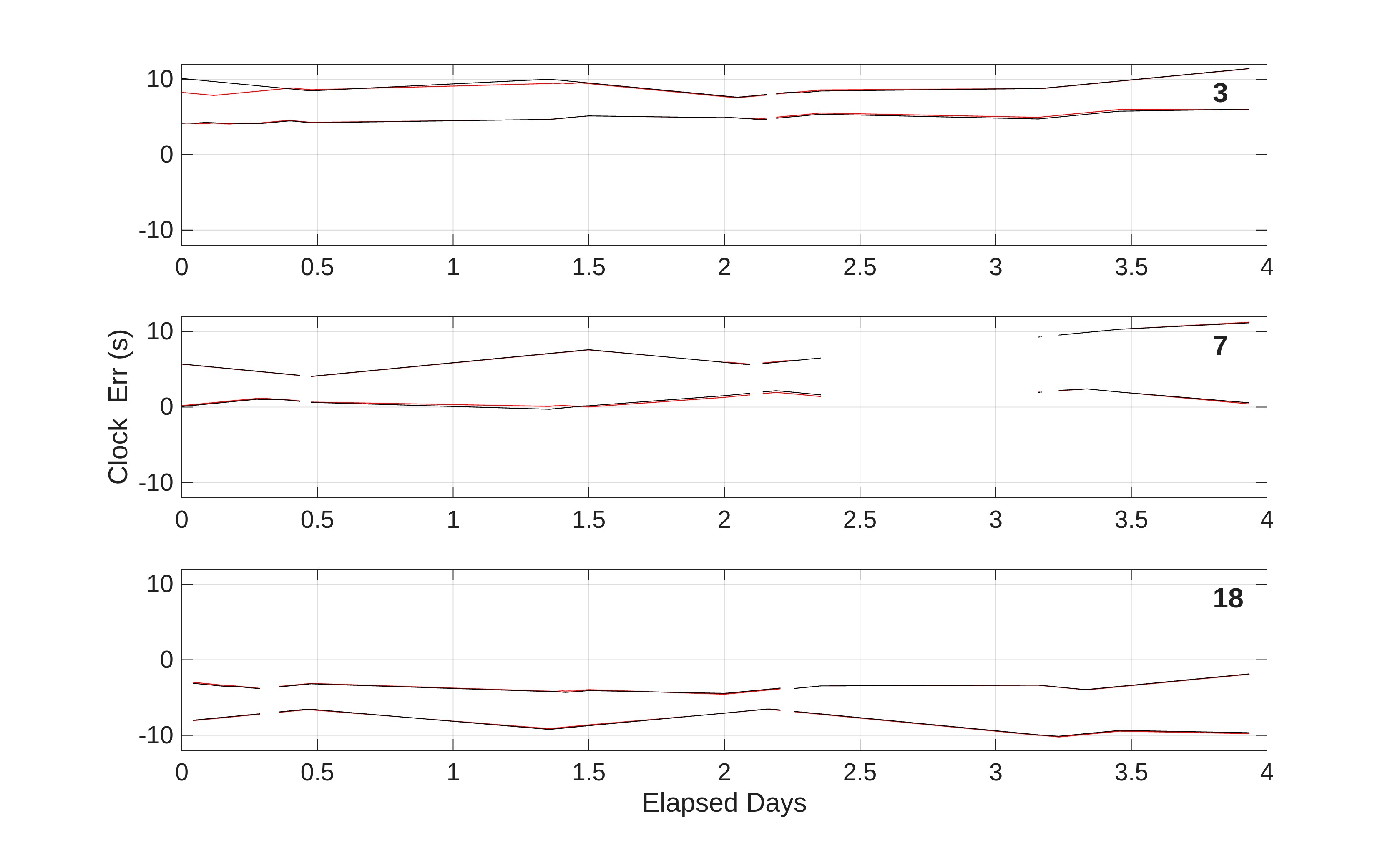}}
\caption{\label{fig:clk_bnds}  Lower and upper 100\% confidence intervals  for clock errors with respect to receiver 15 derived with SBE derived using
  four (black) and five (red) receivers. Receiver numbers in upper right (Fig. \ref{fig:bathy_rec}). Positive means clock is fast compared
  with receiver 15.  Gaps indicate no TDOA data.}
\end{figure}

The  posterior bounds of the relative clock errors are 3 to 10 s wide (Fig. \ref{fig:clk_bnds}); an enormous reduction compared with their prior
bounds ($\pm 2 \times 10^4$ s; Sec. \ref{sec:clocks}).
Prior and posterior bounds of the receiver locations are the same.
Posterior bounds of the $c_{3d}$ occasionally show a slight contraction compared with prior values (not shown).

\begin{figure}[ht!] 
   \centerline{\includegraphics[width=4.in]{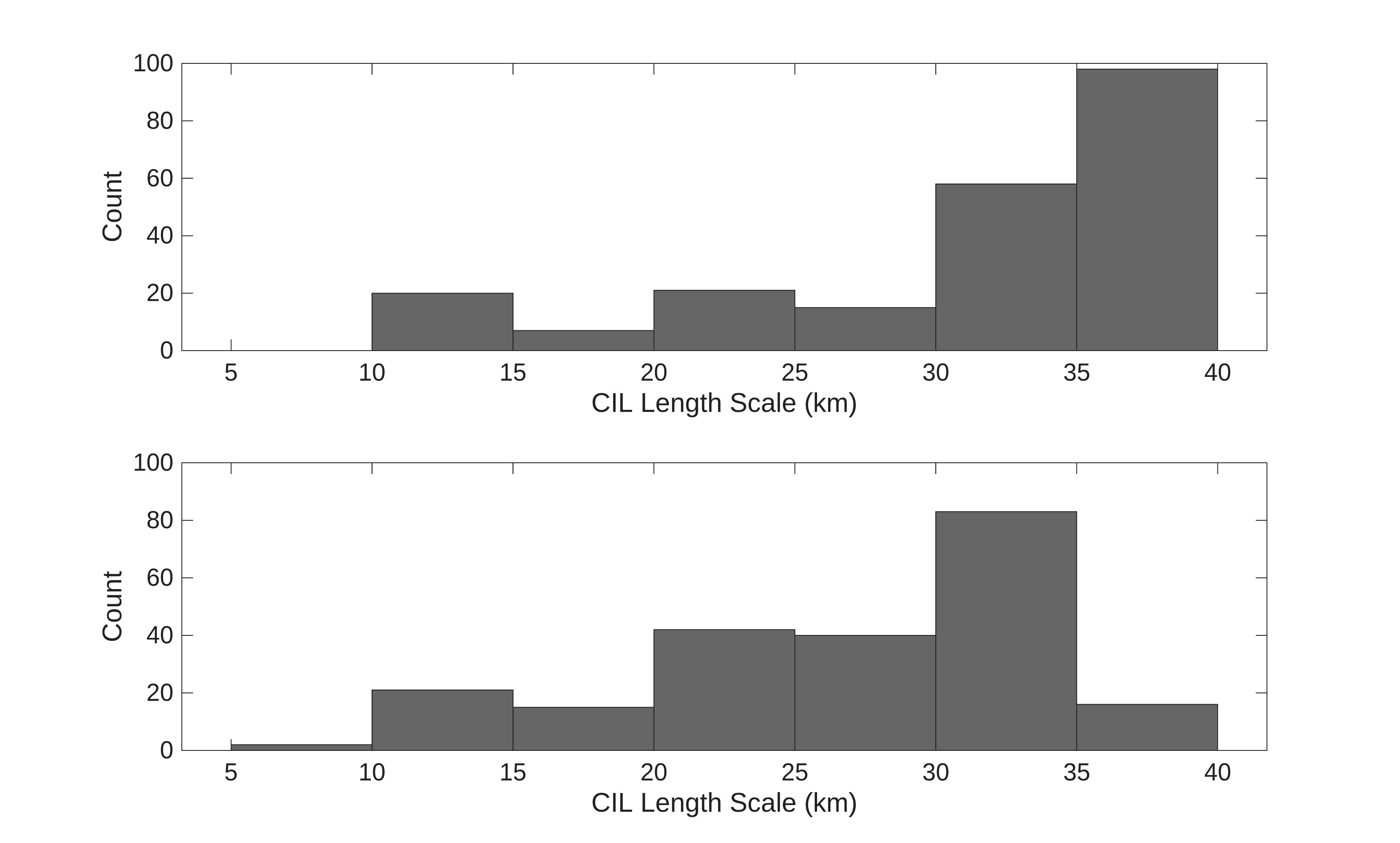}}
\caption{\label{fig:histo_len_scales}  Length scales of CIL derived with SBE with four (top) and five (bottom) receivers. Length scale is square root of CIL.  Minimum  length scales are 11.3 and 9.7 km respectively.
}
\end{figure}

The CIL for the explosions should and do decrease with the number of receivers (Fig. \ref{fig:histo_len_scales}).  The decreases are about 8\% and 4\% for explosions
near and far from the receivers (not shown), where the length scale is the square root of the CIL. The top row of Fig.  \ref{fig:fig_for_article_draft} shows
data from two explosions when five receivers yield
much smaller CIL. The bottom row shows two other explosions where the CIL only contract a small amount. The four and five  receiver results are derived from
the same 219 explosions.

\section{\label{sec:discussion} Discussion}

\subsection{\label{sec:prob_correct} SBE is not  Lucky}

The prior bounds for the explosion's locations, [-70.8 -70.4] $\mbox{}^{\circ}$E and [40.14 to 40.55] $\mbox{}^{\circ}$N (Sec. \ref{sec:sus_prior_locs})
encompass 33.9 km of longitude  and 65.6 km of latitude.
This area is $a_{\mbox{prior}} = 2.22 \times 10^{3} \ \mbox{km}^2$. Let $a_i$ be the area of the $i$'th 100\% CIL derived from  SBE.
If the CIL have no predictive ability, i.e. their boundaries are uniformly distributed in the region,
the probability the true location is in the $i$'th CIL is $p_i = a_i/a_{\mbox{prior}}$. The probability all 219 GPS locations are in their respective
CIL by luck is $P = \prod_{i=1}^{219} p_i$. Evaluating $a_i$ from the five receiver dataset yields $P=4.3  \times 10^{-116}$.  Therefore, SBE's
successful containment of all 219 explosions is not due to luck.

%
\subsection{\label{sec:sim1} Effects of Clock Errors on the CIL}

We investigate if the large CIL from the  experiment (Fig. \ref{fig:histo_len_scales}) are inherently due to the large clock errors.
To  this end, two   simulations of the experiment are conducted, where the locations of fifty explosions are randomly selected within
the prior bounds of the experiment. The simulations adopt a TDOA uncertainty of $\pm 0.01$ s, instead of the $\pm 0.5$ s due to
A/D  timing problems across soundtrap audio files (Sec. \ref{sec:tdoa_bnds}).
An uncertainty of 0.01 s is achievable by cross-correlating time series  with an root-mean-square bandwidth of 5 Hz and a
SNR of 20 dB (Eq. 3.17, p. 278 \cite{helstrom2013statistical}).  Larger bandwidths and SNRs yield smaller uncertainty.

\begin{figure}[ht!] 
  \centerline{\includegraphics[width=3.5in]{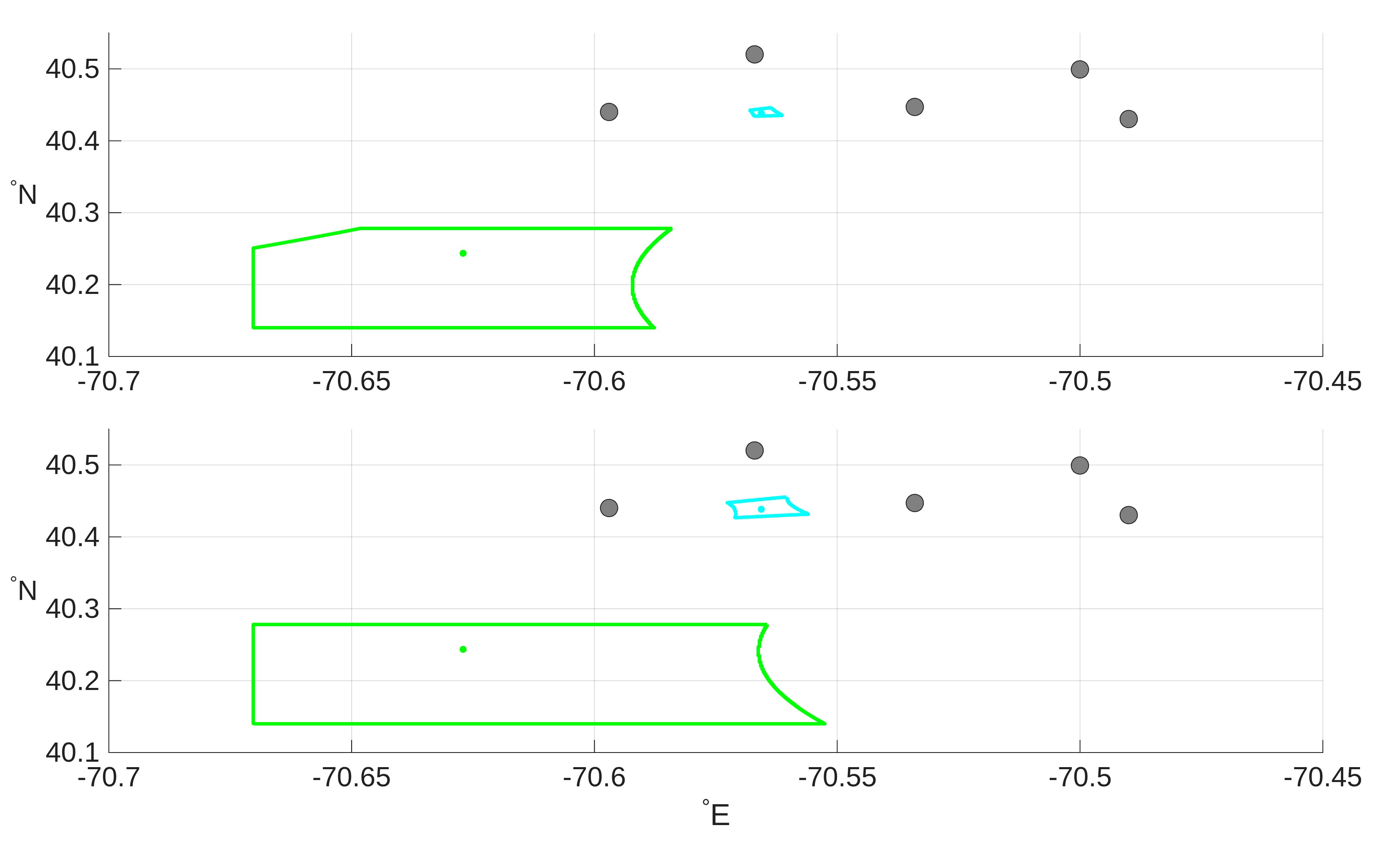}}
\caption{\label{fig:simulation1_and_2} {\bf Top}: Simulation of SBE's 100\% CIL for two  sounds (green and cyan) with true locations shown
  and same five receivers from experiment but clocks are synchronized
  and uncertainty of each TDOA reduced from $\pm 0.5$ s to $\pm 0.02$ s. Bounds of $c_{3d}$  are [1458 , 1515] m/s, similar to
  those used for the data when source is not close to a receiver (Fig. \ref{fig:c3d_bnds_T03}). {\bf Bottom}: Same except clock errors are same
as experiment assuming there are no $\pm 0.5$ s A/D time stamp discontinuities across audio file boundaries (Sec. \ref{sec:tdoa_bnds}).}
\end{figure}

The first simulation  assumes clocks are synchronized.   The 100\% CIL are  small when 
explosions are near the receivers and larger otherwise (Top, Fig.  \ref{fig:simulation1_and_2}). 
The second simulation assumes the same large clock  errors as the experiment (Bottom,
Fig.  \ref{fig:simulation1_and_2}). Both simulations yield small CIL when explosions are near  the  receivers.
Thus it should be possible to obtain small CIL with inaccurate clocks. One possible contribution to the simulator's smaller CIL is the
random locations of the explosions. Random locations are more likely effective in synchronizing clocks as can  be understood by studying the
algebraic solutions of clock  error from TDOA data \cite{patent_clock_sync1,patent_clock_sync2}.

We conclude the large  CIL  from the experiment are not inherently due to the large clock errors if there are no $\pm 0.5 s$ discontinuities
in timing across audio file boundaries.
It is  beyond the  scope of this paper to further analyze the
causes of the relatively larger CIL from the experiment.

\subsection{\label{sec:animal_locs} CIL for Whale Calls} 

Marine bioacousticians sometimes locate calling whales from time-unsynchronized receivers
\citep{pass_acous2, report1, clark_charif,  conf_1, conf_2, mennil, BOEM2018, BOEM2019}.  The algorithm employed
may not account for the clock errors, resulting in large errors recently reported \cite{cse_eval,ishmael}. 
The results in this paper  may be of interest to this community, 
so a few remarks
are in order.
The simulation shown at the top of Fig. \ref{fig:simulation1_and_2} predicts the CIL for calling marine mammals may have length scales of a kilometer
when near the receivers and the clocks are synchronized.   When the errors of  the clocks are as large
as the experimental data here, and the calls originate from a  wide variety of spatial  locations, the length scales of their CIL are a few kilometers when
those calls are near the receivers (bottom, Fig. \ref{fig:simulation1_and_2}).  The CIL become large when the calls are far from  the receivers,  regardless
of whether the clocks are synchronized or not.

\subsection{\label{sec:sus_prior} SBE's Reliability Independent of Prior Bounds Size}

SBE's reliability is independent of the sizes of its prior bounds, as long as they contain the true answer \cite{sbe}. There is no proof this is so,
but has been found to be so in hundreds of thousands of simulations and in various assimilations of the data in this experiment (not shown).
The result is sensible because the information conveyed by the TDOA data are independent of the sizes of  the prior bounds. In other words, when
the prior bounds are larger, there are more scenarios to eliminate with the data, though the information from the data are unchanged.
The prior locations of the SUS charges could be much larger because they can be detected hundreds of kilometers away
(Sec. \ref{sec:sus_prior_locs}). The CIL would increase, but the amount of increase is not linear with bound size and the TDOA data will
reduce them as much as the physics allows. The non-linearity of the increase is a consequence of the non-linear relation between the TDOA
and location.

\section{\label{sec:conclusion} Conclusion}

A blind evaluation was conducted of SBE's CIL using sounds from 219 explosions.
Its 100\% CIL contained the explosion's GPS derived locations without exception.
The probability this outcome occurred by chance was $P=4.3  \times 10^{-116}$. 
 We found the CIL decreased when increasing the number of receivers from four to five. It would be valuable to
further evaluate SBE's  reliability with many more of the twenty receivers.

This evaluation of SBE's reliability compliments a previous classified evaluation with real-data \cite{sbir_phase2} and
thousands of simulations \cite{cse_eval,ishmael}.
These 100\% CIL also contained the true locations in every case.

\begin{acknowledgments}
  Research  was supported by Office of Naval Research grant\blue{s} N00014-23-1-2336 and N00014-26-1-2087. We thank
  Preston Wilson and David Dall'Osto for deploying the SUS charges  and
  measuring their locations with the GPS. We thank Oleg Godin for identifying analytical expressions for the amplitudes of caustics and
  explaining their implementation.
\end{acknowledgments}

\section{\label{declarations} Author Declarations}
The authors declare no conflicts of interest.

The TOSSIT data used in this study are available upon request to Dr. Bonnel.



\appendix*
\section{Removing clock discontinuities from data}                       
\label{sec:app1}

Data were recorded with OceanInstruments software version 4.0.0.4 and micro-controller firmware version 214 for soundtrap model ST300STD.
This system creates so-called SUD files.   They are converted to wav files with OceanInstruments software version 4.0.21 (10May2024), with
the zero-filled option enabled to fill missing digital  samples.  The log files created by the conversion to audio files stated there were
no missing data.  We informed OceanInstruments of these findings, and they stated the zero fill option may not be working
correctly. 


Each audio file created by the ST300STD outputs an ascii log file tabulating the times of the first and last A/D sample and the
number of samples.  The tabulated start and end times disagree with the number of samples, and we found the number of samples
to be a better estimator of the A/D timestamps crossing from one audio file to the next.
This problem can be  seen by comparing the measured propagation time of a SUS charge with its predicted time based on
a  representative speed of sound and the locations of it and a receiver.

\begin{figure}[ht] 
  \centerline{\includegraphics[width=6in]{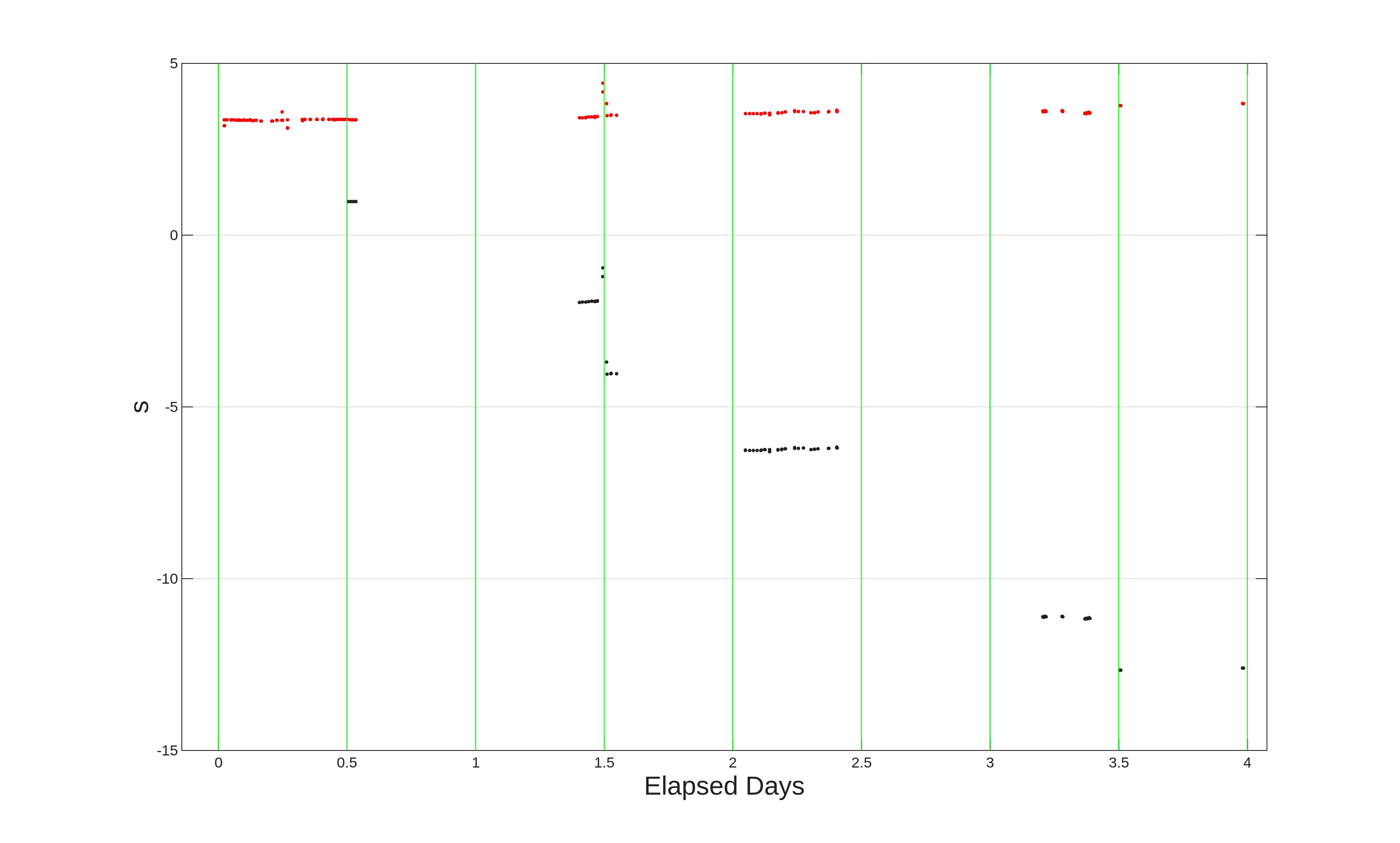}}
\caption{\label{fig:correct_time_jumps} Measured minus predicted times of received explosion times at receiver 3 across nine consecutive
  audio files,  each of about 12 hr duration, over four days (black). Predicted times derived by computing distances between GPS SUS location
  and receiver and dividing by about 1475 m/s. Note change in measured minus modeled across audiofile boundaries (green vertical lines). Red dots are  same except
  corrected for discontinuities using algorithm with  Eq. \ref{eq:t_detect_correction}.  No correction is  made for times for first audiofile;
  thus those red dots
  overlay the black dots.}
\end{figure}

The measured time difference between the $q$'th SUS detonation, $t_{expl}(q)$, and its detected time of arrival at a  receiver, $t_{detect}(q)$,
measured with its imperfect clock, is,
\begin{equation}
T_q \equiv t_{detect}(q) - t_{expl}(q) \ . \label{eq:T_q}
\end{equation}
where $t_{detect}(q)$ is computed from the start time  of the first A/D sample,
tabulated in its log file, and the number of elapsed samples in the audio file up to its detection time.
The detonation  time,  $t_{expl}(q)$, is measured on the ship.
In the absence of clock error, $T_q$ would be the travel time between the q'th explosion and the receiver of interest. Let $\hat{T}_q$ be the modeled value of $T_q$ derived from the horizontal distance between explosion $q$ and the receiver and some approximate value
for  the  speed of sound.
Their difference , $\delta T_q \equiv  T_q-\hat{T}_q$, should not vary much with $q$ if the receiver's clock is perfect. But the receiver's clock can drift many seconds per day,
so $\delta T_q$ should and does change significantly during the experiment, as can be seen in Fig. \ref{fig:correct_time_jumps} (black dots). It is important
here to notice here two type of variation: a slow drift likely associated with small change in the clock and/or propagation conditions, and drastic jump between audio files (with file boundaries indicated by vertical green lines). The jumps between audio files are a critical problems for localization if they are not properly handled.

To correct the detection times, it is important to understand some of the log files. A log file contains lines that look like: \\
\textit{SamplingStartTimeUTC="2022-05-15T19:57:59"} \\
\textit{SamplingStopTimeUTC="2022-05-16T07:57:04''} \\
\textit{WavFileHandler SampleCount ="1035564144''}  \\
For log file number $i$, the first and second lines are supposed to equal the time stamps of the first and last A/D samples in audiofile $i$. The third line is supposed
to be the number of samples inn audio  file $i$. In the following, those variable are respectively defined as $l_{start}(i)$, $l_{stop}(i)$ and $n_s(i)$, with
$l_{start}(i)$ and $l_{stop}(i)$ defined as matlab date number (i.e. a calendar date as the total number of days elapsed since a fixed reference), and $n_s$ an integer.

The number of samples in an audio file and the audio files' start time  yields another estimate of the time  stamp of the last A/D  sample as,
\begin{equation}
 l_{stop}^{(derive)}(i)=l_{start}(i)+ (n_s-1)/f_s/86400
\end{equation}
where $f_s$ the sampling frequency in Hz. The mismatch, in seconds, between the start of file $i+1$ and the end of file $i$ is defined as
\begin{equation}
l_{mismatch}(i+1)=[l_{start}(i+1)- l_{stop}^{(derive)}(i)]/86400 \ ,
\end{equation}
and $l_{mismatch}(1) \equiv 0$.

Form the cumulative sum,
\begin{equation}
\alpha(i) = \sum_{j=0}^{i-1} l_{mismatch}(j+1) \ ,
\end{equation}
and define the vector,
\begin{equation}
  {\bf v}=[0 ~~\alpha(2) ~~\alpha(3) \ \cdots \ \alpha(N)] \ , 
\end{equation}
where there are $N$  audio files.

Define,
\[
ind_{finite}(i) =
\begin{cases}
    1 &  \ , \ \mbox{when at least one  detected explosion in  audiofile number $i$}\\
    0 & \ , \ \mbox{otherwise .}
    \end{cases}
\]
For each $i$  assign,
\[
t_{correct}(i) =
\begin{cases}
  0 & \ , \ \text{if}~~i=1\\
  -v(i)  \times ind_{finite}(i) & \ , \ \text{otherwise.} 
\end{cases}
\]
Note $t_{correct}(i)$ is a scalar.

Discontinuities due to bugs in the soundtrap software/hardware are corrected for audiofile number $k$ using,
\begin{equation}
t_{detect~correction}(k)=t_{detect}(k) + t_{correct}(k). \label{eq:t_detect_correction} 
\end{equation}

\begin{figure}[ht] 
  \centerline{\includegraphics[width=6in]{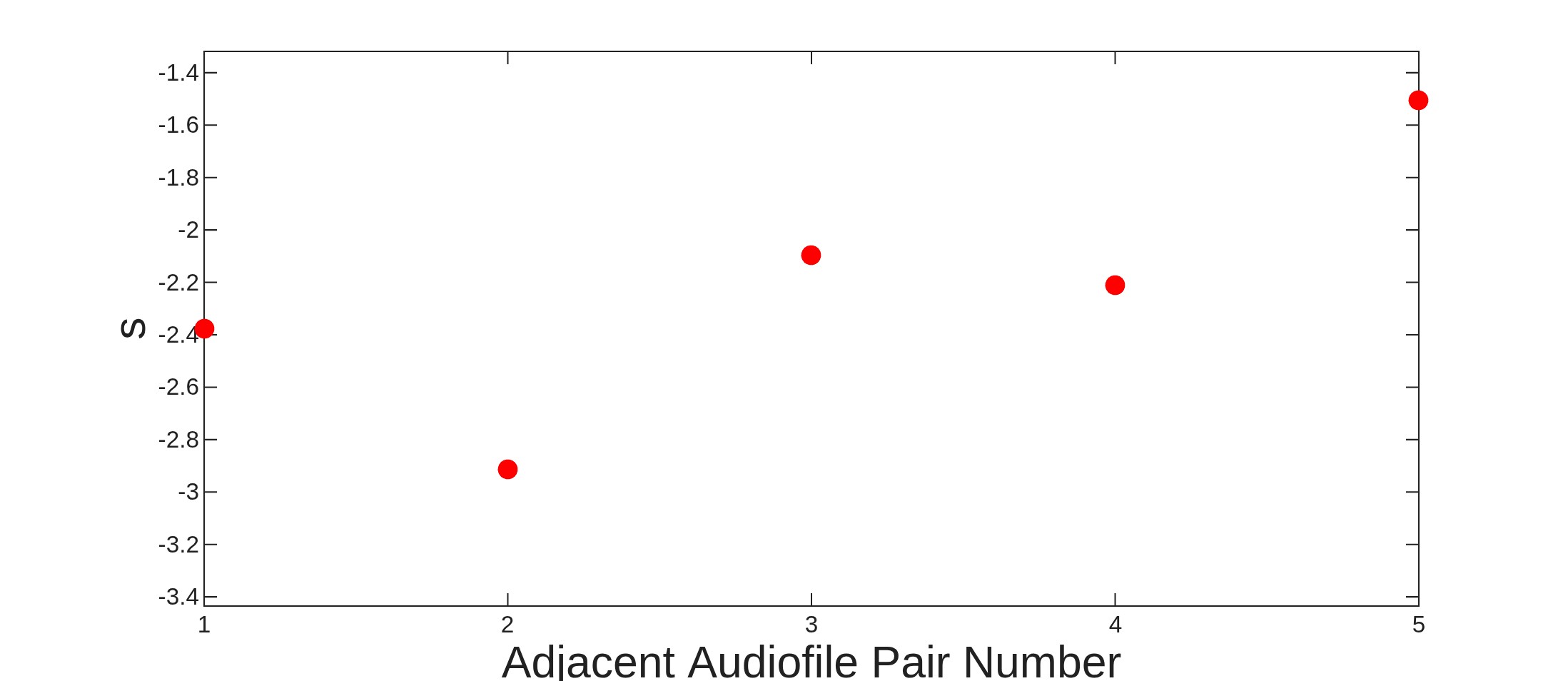}}
\caption{\label{fig:time_jumps}  Discontinuities of measured minus modeled (MMM) SUS charge times at receiver 3. The measured time of detection
  from an audiofile  is the median of all  detection times.  Six MMM values plotted (red) correspond to median times between audio files 1 and 2, 3 and 4, 4 and 5, and 7 and 8 corresponding
  to pair numbers 1 through 6 on x axis. There are no
measured data in audio files 6 and 9 for receiver 3.}
\end{figure}

If $t_{detect~correction}(q)$ is used in place of $t_{detect}(q)$ in Eq. \eqref{eq:T_q},
the large jumps across audiofile boundaries are much reduced (red dots, Fig. \ref{fig:correct_time_jumps}).
Fig. \ref{fig:time_jumps} shows discontinuities of the median value of $t_{detect}(q)$, for each audiofile, from receiver 3.
These are removed using Eq. \eqref{eq:t_detect_correction}. 

OceanInstruments informed us of another possible clock error of $\pm 0.5$ s across boundaries of the audio files.
It might be possible to see if this occurs by more closely examining differences between modeled and measured propagation times on either
side of the audiofile boundaries.
This has not been done for this paper.

\clearpage





\bibliography{sampbib}


\end{document}